\documentstyle[12pt]{article}
\begin{document}
\def\thefootnote{\fnsymbol{footnote}}
\begin{flushright}
KANAZAWA-03-30  \\ 
October, 2003
\end{flushright}
\vspace*{2cm}
\begin{center}
{\LARGE\bf  Gaugino CP phases and EDMs 
in the extended gauge mediation SUSY breaking}\\
\vspace{1 cm}
{\Large Daijiro Suematsu}
\footnote[1]{e-mail: suematsu@hep.s.kanazawa-u.ac.jp}
{\Large and Hirokazu Tsuchida}
\footnote[2]{e-mail: tsuchida@hep.s.kanazawa-u.ac.jp}
\vspace*{1cm}\\
{\it Institute for Theoretical Physics, Kanazawa University,\\
        Kanazawa 920-1192, Japan}\\    
\end{center}
\vspace{1cm}
{\Large\bf Abstract}\\  
We study phenomenological aspects of the soft supersymmetry 
breaking parameters in a model with the extended gauge mediation 
supersymmetry breaking. In this model
gaugino masses can be non-universal and as its result 
physical CP-phases remain in the gaugino sector even after the 
$R$-transformation. 
These phases contribute to the electric dipole moment (EDM) of an electron
and a neutron. We show that their experimental bounds can be 
satisfied even for the situation such that there exist the order 
one CP-phases and the masses of superpartners are of the order of 100~GeV.
\newpage
\setcounter{footnote}{0}
\def\thefootnote{\arabic{footnote}}
\section{Introduction}
At present low energy supersymmetry seems to be the most promising candidate
for a solution of the gauge hierarchy problem or the weak scale
stability. 
Although we have no direct evidence for the supersymmetry still now, 
the gauge coupling unification found in the minimal supersymmetric 
standard model (MSSM) may be considered as its indirect signal.
When we consider the supersymmetric models, the supersymmetry breaking
mechanism is crucial for their phenomenology. In fact, the experimental
bounds on the flavor changing neutral currents (FCNC) severely 
restrict the supersymmetry breaking in the observable sector.
They require the masses of the scalar superpartners to degenerate
strictly. The gauge mediation supersymmetry breaking (GMSB)
\cite{gm,mgm1,mgm2,exmgm1,exmgm2,gmsbrev} is promising from 
this point of view because of its flavor blindness.

Another phenomenological constraint on the supersymmetry breaking 
comes from the electric dipole moment (EDM) of an electron and a neutron. 
It is well-known that the EDM of the electron and the neutron 
should be severely suppressed on the basis of the 
experimental data \cite{edmb}. 
It has been recognized that there are two possibilities to satisfy this
constraint \cite{edmt}. One is that the soft breaking parameters are taken to 
be of the order of 100~GeV assuming that the soft CP-phases are smaller
than $10^{-2}$. Such small phases are usually considered to 
be unnatural and this aspect is considered as a default of the
supersymmetric models.
The other one is that the soft CP-phases are supposed to be of the
order unity by assuming the soft scalar masses to take larger
values than 1~TeV, which is considered to be 
unattractive from the view point of the weak scale supersymmetry.
If we can have the third possibility in which the order one CP-phases 
and the superpartners with the masses of the order of 100~GeV can be consistent
with the EDM constraints, it is very interesting and we can have a lot 
of interesting phenomenology \cite{cancel1,cancel2,gphase,flav,cp}.
In particular, if we consider the origin of the baryon number asymmetry 
in the universe, we may need new sources of CP violation.
It is known that the Cabibbo-Kobayashi-Maskawa (CKM) phase in the 
standard model (SM) is 
insufficient to explain the baryon number asymmetry through the electroweak
baryogenesis scenario because of a suppression due to the smallness of
the quark 
flavor mixing. The possibility of the order one CP-phases in the soft
supersymmetry breaking parameters seems to be fascinating since such 
soft CP-phases 
present us the promising sources for the CP-phases required 
in the electroweak baryogenesis and also they may allow us to relax the
required Higgs mass bound \cite{Higgs}.

When we consider this third possibility, it is useful to note that
the usual analyses of the EDM are based on the assumption of the universal
gaugino masses as stressed in \cite{gphase}. 
If we loose this assumption, we can find the way out of the ordinary
understanding. In fact, there are interesting suggestions that the 
experimental constraints on the EDM can be satisfied due to the 
effective cancellation among various contributions 
to the EDM\footnote{
In the case of the electron EDM the cancellation between the chargino and
neutralino contributions have been shown to occur 
\cite{cancel1,cancel2,gphase}. 
On the other hand, in the case of the neutron EDM it has been known that 
there are several types of cancellation , that is, the cancellation
between the gluino and the chargino exchange diagrams and also the 
cancellation among the gluino exchange diagrams in themselves 
{\it etc} \cite{cancel1,gluino}. The combined effect of these 
cancellations allows the possibility of 
the large soft CP-phases \cite{cancel1,cancel2,gphase}. }
as far as the CP-phases in the gaugino masses are non-universal
even in the case where both the order one CP-phases and rather light
superpartners exist. 
Unfortunately, the non-universal gaugino masses seems to be 
rather difficult to be realized 
in both the unified theory and the superstring as discussed 
in \cite{gphase,univ}.
However, if we consider the GMSB such that some kind of discrete 
symmetry imposes the SU(3) triplet and SU(2) doublet messenger fields 
couple to the different 
singlet fields where the supersymmetry is broken due to the hidden 
sector dynamics, we can show that
the non-universal phases in the gaugino masses 
appear naturally \cite{sue}.
In the following discussion we will assume this extended structure in
the messenger superpotential.  
In such a model the naturalness problem for the soft CP-phases may 
disappear since the EDM bounds for the electron and the neutron are 
satisfied by the
cancellation among various contributions even for the order one CP-phases.
Since the soft supersymmetry breaking parameters induced by the 
GMSB scenario is strongly constrained, we can survey such a possibility 
in the wide parameter region without large ambiguity.

This paper is organized as follows. In section 2 we discuss 
the soft supersymmetry breaking parameters in the extended GMSB.
In section 3 we briefly review the formulas of the EDM of the electron and 
the neutron.
In section 4 we discuss the feature of the soft supersymmetry 
breaking parameters 
using the numerical analysis based on the renormalization group equations.
We also present our result for the estimation of the EDM in 
this model. The anomalous magnetic moment of a muon is predicted for
these parameters. Section 5 is devoted to the summary.
In the appendix we present an example for the realization of the 
required messenger superpotential. 

\section{A model with the extended gauge mediation}

In this section we introduce the extended GMSB and
present the formulas for the supersymmetry breaking parameters in such a 
model. The messenger sector of the ordinary minimal GMSB model is defined by
\begin{equation}
W_{\rm min}=\lambda_q Sq\bar q +\lambda_\ell S\ell\bar\ell,
\end{equation}
where the messenger fields $q~(\bar q)$ and $\ell~(\bar\ell)$ are the ${\bf
3}(\bar {\bf 3})$ of SU(3) and the ${\bf 2}(\bar {\bf 2})$ of SU(2),
respectively. It is also assumed that $(q,~\ell)$ can be embedded into the 
${\bf 5}$ of the grand unified group SU(5). 
If both of the scalar component $S$ and the auxiliary $F$ component 
of the gauge singlet superfield $S$ get the vacuum expectation values 
$\langle S\rangle$ and $\langle F_S\rangle$ due
to a suitable dynamics in the hidden sector, the gaugino masses and the 
soft scalar masses are generated at one-loop and two-loop level,
respectively.  
If $\lambda_{q,\ell}^2\langle S\rangle^2 \gg \langle F_S\rangle$ is
satisfied, their formulas are known to be the following simple forms
by using $\Lambda=\langle F_S\rangle/\langle S\rangle$,
\begin{eqnarray}
&&M_r=c_r{\alpha_r\over 4\pi}\Lambda, \quad 
\alpha_r={g^2_r \over 4\pi}, \quad 
c_3=c_2={3\over 5}c_1=1, 
\nonumber\\
&&\tilde m_f^2=2\vert\Lambda\vert^2
\left[C_3\left({\alpha_3\over 4\pi}\right)^2 
+C_2\left({\alpha_2\over 4\pi}\right)^2 +{5\over 3}
\left({Y\over 2}\right)^2\left({\alpha_1\over 4\pi}\right)^2\right], 
\end{eqnarray} 
where $C_3=4/3$ and 0 for the SU(3) triplet and singlet fields, and
$C_2=3/4$ and 0 for the SU(2) doublet and singlet fields, respectively. 
The hypercharge $Y$ is expressed as $Y=2(Q-T_3)$. 
The soft supersymmetry breaking parameters $A_f$ and $B$ for the scalar
trilinear and bilinear terms are model dependent and cannot be directly
related to the above formulas. If we take account of  the effects of the 
radiative corrections, they can be written as \cite{exmgm2}
\begin{eqnarray}
&&A_f\simeq A_f(\Lambda)+M_2(\Lambda)\left(-1.85+0.34\vert h_t\vert^2\right)
+\cdots ,
\nonumber\\
&&{B\over\mu}\simeq {B\over\mu}(\Lambda)-A_t(\Lambda)
+M_2(\Lambda)\left(-0.12+0.17\vert h_t\vert^2\right)+\cdots,
\label{eqg}
\end{eqnarray}
where $A_f(\Lambda)$ and $B(\Lambda)$ are the initial values at which
the supersymmetry breaking is introduced.
In the expression of $A_f$ the term with the top Yukawa coupling $h_t$ 
should be neglected except for the top quark sector.

From these formulas we find that there cannot remain the physical
CP-phases in the gaugino sector. In fact, even if the $\Lambda$ is a
complex, they can be rotated away by the $R$-transformation. Thus the
physical CP-phases in the supersymmetry breaking parameters 
are confined in $A_f$ and $B$.
In the case of $A_f(\Lambda)=B(\Lambda)=0$ which may be expected in many
GMSB scenario, $A_f$ and $B$ are proportional to 
the gaugino mass and then the CP-phases in the soft supersymmetry breaking
parameters are completely rotated away
\cite{mgm2,exmgm2}.\footnote{Although this is an interesting solution for 
the soft CP phase problem at least in the case of the real $\mu$, we do
not take this possibility here.}

Now we consider to modify the superpotential $W_{\rm min}$ for 
the messenger fields. We assume that $(q,~\bar q)$ and
$(\ell,~\bar\ell)$ couple with the different singlet chiral superfields 
$S_{1,2}$ due to some kind of symmetry \cite{sue}. Then the messenger
superpotential takes the form such as
\begin{equation}
W_{\rm ext}=\lambda_q S_1q\bar q +\lambda_\ell S_2\ell\bar\ell.
\end{equation}
If the singlet fields $S_1$ and $S_2$ couple with the 
hidden sector fields where the supersymmetry breaks down,
$q, \bar q$ and $\ell, \bar\ell$ play the role of messenger fields as
in the ordinary scenario.
Only difference from the ordinary minimal GMSB scenario is that 
in the superpotential $W_{\rm ext}$ $q,~\bar q$ and $\ell,~\bar\ell$ 
couple with the different singlet chiral superfields.
If we assume that both $S_\alpha$ and $F_{S_\alpha}$ get the VEVs due to 
the couplings with the supersymmetry breaking 
sector, the gaugino masses and the soft scalar masses are 
generated at one-loop and two-loop level, respectively, in the same way as 
the above mentioned ordinary case.
However, the mass formulas are somewhat modified from the usual ones
since the messenger fields $(q,~\bar q)$ and $(\ell,~\bar\ell)$ 
couple with the different singlets. 
 
The gaugino masses can be written in the form as \cite{sue}
\begin{equation}
M_3={\alpha_3\over 4\pi}\Lambda_1, \qquad
M_2={\alpha_2\over 4\pi}\Lambda_2, \qquad
M_1={\alpha_1\over 4\pi}\left({2\over 3}\Lambda_1+\Lambda_2\right). 
\label{eqff}
\end{equation}
It is interesting that these formulas show that $M_3$ can be smaller 
than $M_{1,2}$ in the case of $\vert\Lambda_2\vert>\vert\Lambda_1\vert$. 
If we take account of the renormalization group evolution effect,
their values at the weak scale $M_W$, for example, can be obtained as
\begin{equation}
M_r(M_W)=M_r(\Lambda){\alpha_r(M_W)\over\alpha_r(\Lambda)},
\end{equation}
where $\Lambda$ is a scale at which the supersymmetry breaking is introduced.
Since $\Lambda_\alpha$ is generally independent, the phases contained in the 
gaugino masses are non-universal even in the case of 
$\vert\Lambda_1\vert=\vert\Lambda_2\vert$. 
In that case we cannot remove them completely 
by using the $R$-transformation unlike the case of the 
universal gaugino mass.
In fact, if we define the phases as
$\Lambda_\alpha\equiv\vert\Lambda_\alpha\vert e^{i\theta_\alpha}$ and make
$M_2$ real by the $R$-transformation, the phases of the gaugino masses 
$M_r$ are written as
\begin{eqnarray}
&&\varphi_3\equiv{\rm arg}(M_3)=\theta_1-\theta_2, \qquad 
\varphi_2\equiv{\rm arg}(M_2)=0, \nonumber \\ 
&&\varphi_1\equiv{\rm arg}(M_1)=\arctan\left({2\vert\Lambda_1\vert
\sin(\theta_1-\theta_2)\over
3\vert\Lambda_2\vert+2\vert\Lambda_1\vert\cos(\theta_1-\theta_2)}\right).
\label{eqfff}
\end{eqnarray}
The scalar masses are induced through the two-loop diagrams as in 
the ordinary case. Their formulas can be written as
\begin{equation}
\tilde m^2_f=2\vert\Lambda_1\vert^2
\left[C_3\left({\alpha_3\over 4\pi}\right)^2 
+{2\over 3}\left({Y\over 2}\right)^2\left({\alpha_1\over 4\pi}\right)^2\right]
+2\vert\Lambda_2\vert^2
\left[C_2\left({\alpha_2\over 4\pi}\right)^2 
+\left({Y\over 2}\right)^2\left({\alpha_1\over 4\pi}\right)^2\right].
\label{eqe}
\end{equation}
If $\vert\Lambda_2\vert>\vert\Lambda_1\vert$ is realized, 
the SU(2) doublet fields tend to be
heavy. As this result the color singlet fields can be heavier than the
colored fields depending on the values of $\Lambda_{1,2}$. 
This seems to be a large difference from the ordinary
scenario.
As in the minimal GMSB model the soft supersymmetry breaking $A_f$ 
and $B$ parameters are model dependent also in this case.
However, even in the case of $A_f(\Lambda)=B(\Lambda)=0$ 
there can remain the physical CP-phases in the gaugino sector 
since the phases in the gaugino masses are not universal in general.

\section{EDM and AMM of the leptons}
We consider the MSSM with the soft supersymmetry breaking parameters 
which can be expressed by the mass formulas presented in the previous section. 
At first, we briefly review a relevant part of the MSSM to the study
of the EDM of the quarks and leptons in order to fix the notation used here.
Superpotential related to the lepton sector is given as
\begin{equation}
W=\sum_j \left(h_j^U Q_j H_2\bar U_j +h_j^D Q_j H_1\bar D_j+
h_j^L L_j H_1\bar E_j \right)+ \mu H_1H_2,
\label{eqh}
\end{equation}
where we take the basis in which the flavor mixings are resolved.\footnote{
We do not consider a Yukawa coupling for neutrinos, for simplicity.}
A supersymmetric mass parameter $\mu$ can be complex.
The relevant soft supersymmetry breaking terms are introduced as
\begin{eqnarray}
-{\cal L}_{\rm soft}&=&\sum_\alpha \tilde m_\alpha^2\vert\phi_\alpha\vert^2
+\left\{ \sum_j \left(
A_j^U h_j^U\tilde Q_j H_2\tilde{\bar U_j}
+A_j^D h_j^D\tilde Q_j H_1\tilde{\bar D_j}
+A_j^L h_j^L\tilde L_j H_1\tilde{\bar E_j}\right)\right. \nonumber \\ 
&+&\left.B\mu H_1H_2
+{1\over 2}\sum_r M_r\lambda_r\lambda_r +{\rm h.c.}\right\}, 
\label{eqi}
\end{eqnarray}
where we put a tilde for superpartners of the chiral superfields
corresponding to the SM contents. 
The first term represents soft supersymmetry breaking masses for
all scalar components of the chiral superfields. The third term in the
parentheses represents the
gaugino mass terms. Soft parameters $B$ and $A_j$ are the 
coefficients of the bilinear and trilinear scalar couplings 
and have a mass dimension.
Although soft supersymmetry breaking parameters $A_j$, $B$ and
$M_r$ can generally include the CP-phases, all of these are not
independent physical phases. If we use the $R$-symmetry and redefine the 
fields appropriately, we can select out the physical CP-phases among them.
We take them as
\begin{equation}
A_j=\vert A_j\vert e^{i\phi_{A_j}},\quad
\mu=\vert \mu\vert e^{i\phi_\mu}, \quad
M_r=\vert M_r\vert e^{i\phi_r} ~~ (r=1,3),
\label{eqii}
\end{equation}
where $B\mu$ and $M_2$ are real.
These effective CP-phases are related to the original phases $\varphi_i$ 
in the complex parameters introduced in eq.~(\ref{eqfff}) as follows,
\begin{equation}
\phi_{A_j}=\varphi_{A_j}-\varphi_2, \quad
\phi_\mu=-\varphi_B +\varphi_2, \quad
\phi_{1,3}=\varphi_{1,3}-\varphi_2.
\end{equation}
It should be noted that in this definition
the VEVs of the doublet Higgs scalars 
$H_1$ and $H_2$ are taken to be real.

The mixing matrices in the sleptons, charginos and neutralinos are 
important elements to write down the formula for the EDM 
at the one-loop approximation. 
The mass terms of charginos can be written as
\begin{equation}
-\left(\tilde H_2^+, -i\lambda^+\right)
\left(\begin{array}{cc}\vert\mu\vert e^{i\phi_\mu}& \sqrt 2m_Zc_W\sin\beta\\
\sqrt 2m_Zc_W\cos\beta& M_2 \\ \end{array}\right)
\left(\begin{array}{c}\tilde H_1^- \\ -i\lambda^-\\ \end{array}\right),
\label{eqj}
\end{equation}
where $\tan\beta=\langle H_2\rangle/\langle H_1\rangle$ and 
the abbreviations such as $s_W=\sin\theta_W$ and $c_W=\cos\theta_W$ are used.
The mass eigenstates $\chi_i^\pm$ are defined in terms of the 
weak interaction eigenstates through the unitary transformations 
in such a way as
\begin{equation}
\left(\begin{array}{c}\chi_1^+\\ \chi_2^+\\ \end{array}\right)
\equiv W^{(+)\dagger}\left(\begin{array}{c}\tilde H_2^+\\ -i\lambda^+\\
\end{array}\right),  \qquad
\left(\begin{array}{c} \chi_1^-\\ \chi_2^-\\ \end{array}\right)
\equiv W^{(-)\dagger}
\left(\begin{array}{c}\tilde H_1^-\\ -i\lambda^-\\ \end{array}\right).
\label{eqk}
\end{equation}
Since we consider the GMSB and then the flavor mixing in the sfermion
sector can be neglected, the sfermion mass matrices can be reduced into 
the $2\times 2$ form for each flavor.
This $2\times 2$ sfermion mass matrix can be written in terms of 
the basis $(\tilde f_{L_\alpha}, \tilde f_{R_\alpha})$ as
\small
\begin{equation}
\left(\begin{array}{cc} |m_\alpha|^2
+\tilde m_{L_\alpha}^2+ D_{L_\alpha}^2& 
m_\alpha(\vert A_\alpha\vert e^{i \phi_{A_\alpha}}
+\vert\mu\vert e^{-i\phi_\mu} R_f)\\
m_\alpha(\vert A_\alpha\vert e^{-i\phi_{A_\alpha}}+
\vert\mu\vert e^{i\phi_\mu} R_f)& 
|m_\alpha|^2+\tilde m_{R_\alpha}^2+ D_{R_\alpha}^2\\ 
\end{array}\right),
\label{eql}
\end{equation}
\normalsize
where $m_\alpha$ and $\tilde m_{L_\alpha,R_\alpha}^2$ are the masses 
of the ordinary fermion 
$f_\alpha$ and its superpartners $\tilde f_{L_\alpha,R_\alpha}$, respectively. 
$R_f$ is $\cot\beta $ for the up component of the SU(2) fundamental
representation and $\tan\beta$ for the down component. 
$D_{L_\alpha}^2$ and $D_{R_\alpha}^2$ represent 
the $D$-term contributions, which are 
expressed as follows,
\begin{eqnarray}
&&D_{L_\alpha}^2=m_Z^2\cos 2\beta(T_f^3-Q_fs_W^2), 
\nonumber \\
&&D_{R_\alpha}^2=m_Z^2s_W^2Q_f\cos 2\beta, 
\label{eqm}
\end{eqnarray}
where $T_f^3$ takes 1/2 for the sfermions in the up sector and $-1/2$ for
the ones in the down sector. $Q_f$ is an electric charge of the field $f$.
We define the mass eigenstates $(\tilde f_1, \tilde f_2)$ by the unitary 
transformation such as
\begin{equation}
\left(\begin{array}{c} \tilde f_1 \\ \tilde f_2\\ \end{array}\right)
\equiv V^{\ell\dagger}\left(\begin{array}{c}\tilde f_L \\ \tilde f_R\\ 
\end{array}\right).
\label{eqn}
\end{equation}

If we take the canonically normalized neutralino basis as 
${\cal N}^T=(-i\lambda_1, -i\lambda_2, 
\tilde H_1^0, \tilde H_2^0)$ and define their mass terms in such a form as
${\cal L}_{\rm mass}^{\rm n}=-{1\over 2}{\cal N}^T{\cal MN}+{\rm h.c.}$,
the 4 $\times$ 4 neutralino mass matrix ${\cal M}$ can be expressed as
\begin{equation}
\left( \begin{array}{cccc}
\vert M_1\vert e^{i\phi_1} & 0 &-m_Zs_W\cos\beta & m_Zs_W\sin\beta \\
0 & M_2 & m_Zc_W\cos\beta & -m_Zc_W\sin\beta \\
-m_Zs_W\cos\beta & m_Zc_W\cos\beta & 0 & -\vert\mu\vert e^{i\phi_\mu} \\
m_Zs_W\sin\beta & -m_Zc_W\sin\beta & -\vert\mu\vert e^{i\phi_\mu} & 0 \\
\end{array} \right).
\label{eqo}
\end{equation}
Mass eigenstates $\chi^0$ of this mass matrix are related to ${\cal N}$ as
\begin{equation}
\chi^0\equiv U^T{\cal N},
\label{eqp}
\end{equation}
where the mass eigenvalues are defined to be real and positive so that 
$U$ includes Majorana phases.

Using these notations we give the formula for the EDM of the charged
leptons. The effective interaction term representing the EDM
of the charged lepton $\ell$ can be written as 
\begin{equation}
{\cal L}_{\rm eff}={1\over 2}~{\cal G}~\bar 
\ell \sigma_{\mu\nu}\ell~ F^{\mu\nu}. 
\label{eqpp}
\end{equation}
The value of the EDM of $\ell$ is related to this effective 
coupling ${\cal G}$ through the formula
\begin{equation}
d_\ell={\rm Im}({\cal G}).
\end{equation}
In the MSSM there are new contributions to $d_\ell$, 
which come from the one-loop diagram with the superpartners of the SM
fields in the internal lines as is well-known \cite{edmt,cancel1,cancel2}. 
These new contributions can be calculated as
\begin{equation}
d_\ell/e=-{\alpha\over 8\pi s^2_W}m_\ell
\left({1\over m_\ell}\sum_{j,a}{1\over m_j}G(x_{aj}) {\rm Im}(A_{\chi^0_j})
+{1\over m_W}\sum_j{1\over m_j} F(x_{\tilde\nu j}){\rm Im}
(A_{\chi^\pm_j})\right),
\label{eqq}
\end{equation}
where $x_{aj}$ is defined as $x_{aj}=\tilde m_a^2/m_j^2$. $m_j^2$ is a mass
eigenvalue of the chargino $\chi^\pm_j$ or the neutralino $\chi_j^0$ 
and $\tilde m_a^2$ is a mass eigenvalue of the slepton $\tilde f_a$.
In the right-hand side of eq.~(\ref{eqq})
the first term represents the neutralino-charged slepton contribution 
and the second term represents the chargino-sneutrino contribution.
$A_{\chi^0_j}$ and  $A_{\chi^\pm_j}$ which express the mixing factors
appearing at each vertex are defined as
\begin{eqnarray}
A_{\chi^0_j}&=&-\left[\left(U_{1j}^2 t^2_W+
U_{1j}U_{2j}t_W\right)V_{1a}^{\ell\ast} V_{2a}^\ell \right. \nonumber \\
&+&\left.{m_e\over 2m_W\cos\beta}\left\{\left(t_W U_{1j}U_{3j}
+U_{2j}U_{3j}\right)\vert V_{1a}^\ell\vert^2 
-2 t_W U_{1j}U_{3j}\vert V_{2a}^\ell\vert^2\right\}
\right],
\nonumber \\
A_{\chi^\pm_j}&=&{1\over\sqrt 2\cos\beta}W^{(-)}_{1j}W^{(+)}_{2j},
\label{eqr}
\end{eqnarray}
where $t_W=\tan\theta_W$ and we neglect a higher order term with respect 
to the charged lepton mass in ${\cal A}_{\chi_j}^0$.
Since we have no right-handed neutrinos or they are considered 
to be decoupled in this expression, 
the slepton in the chargino contribution is fixed to be the
left-handed sneutrino.
Since the fermions in the external lines are very light compared with the
sleptons in the internal lines, $G(x)$ and $F(x)$ are 
approximately written as
\begin{eqnarray} 
&&F(x)={1-3x\over (1-x)^2} -{2x^2\over (1-x)^3}\ln x,\nonumber \\
&&G(x)={1+x\over (1-x)^2} +{2x\over (1-x)^3}\ln x. 
\label{eqs}
\end{eqnarray}

Here it is useful to note the following points to see the cancellation
between the two contributions to the EDM. 
The chargino contribution is related to
the CP-phase $\phi_\mu$ although the neutralino contribution is caused
by the CP-phases $\phi_1,~\phi_\mu$ and $\phi_{A_\alpha}$. 
Generally the chargino contribution can be larger than the neutralino
one because of the existence of the slepton mixing factor in the
neutralino contribution. In order to make both contributions comparable
 the neutralino mass needs to be much lighter than the one of the
chargino in addition to the existence of the suitable CP phases.

The real part of the same one-loop diagram as the ones for the EDM presents
the anomalous magnetic moment (AMM).
Thus it is also expressed
by using the effective coupling ${\cal G}$ in eq.~(\ref{eqpp}) as
\begin{equation}
a_\ell={2m_\ell\over e}{\rm Re}({\cal G}).
\end{equation}
We can calculate the new contribution to $a_\ell$ due to the superpartner
effects in the same way as the EDM
and the result is given by
\begin{equation}
\delta a_\ell=-{\alpha\over 4\pi s^2_W}m_\ell^2
\left({1\over m_\ell}\sum_{j,a}{1\over m_j}G(x_{aj}) Re(A_{\chi^0_j})
+{1\over m_W}\sum_j {1\over m_j} F(x_{\tilde\nu j}) Re(A_{\chi^\pm_j})\right).
\label{eqqq}
\end{equation}
The value of the AMM of the muon is generally affected by the existence 
of the large CP-phases as it can be seen in eqs.~(\ref{eqr}) and (\ref{eqqq}).
Its value can be largely changed from the ordinary estimation which 
is obtained under the assumption of no CP-phases 
if there are large CP-phases in the soft supersymmetry breaking parameters. 
Thus in our model
the estimation of the AMM of the muon can be an interesting subject 
if the experimental bounds of the EDM of the electron and the neutron 
can be satisfied 
for the large CP-phases in the soft supersymmetry breaking parameters.

If we allow the nontrivial CP-phases in the gaugino masses as shown
in eq.~(\ref{eqii}), the gluino mass can have a large CP-phase also.
It can bring a large contribution to the EDM of the neutron 
in addition to the contributions due to the charginos and 
the neutralinos.
We need to investigate them in order to check the consistency of the
present model. In the estimation of the neutron EDM we use its 
nonrelativistic formula based on the quark EDM such as\footnote{In this
analysis we do not consider the contribution to the neutron EDM from the
chromoelectric and the CP-violating purely gluonic dimension six
operators \cite{cancel1}.} 
\begin{equation}
d_n={1\over 3}\left\{(4d_d^g-d_u^g)+ (4d_d^\chi-d_u^\chi)\right\},
\end{equation}
where $d_u$ and $d_d$ are the EDM of the $u$-quark and the $d$-quark
and the superscripts $g$ and $\chi$ represent the contributions from the
one-loop diagrams containing the gluino internal line and the 
chargino/neutralino internal lines, respectively.  
This value of $d_n$ should be
evolved to the hadronic scale by including the QCD correction and then 
it gives $1.53d_n$ \cite{cancel1}.

The gluino contribution to the quarks can be expressed as
\begin{eqnarray}
&&d_f^g/e={\alpha_s\over 6\pi}{Q_f\over \vert M_3\vert}
\sum_{a=1}^2{\rm Im}({\cal A}_g^{f_a})G(x_a), \nonumber \\
&&{\cal A}_g^{f_a}=V^f_{2a}V^{f\ast}_{1a}e^{i\phi_3},
\end{eqnarray} 
where $x_a=\tilde m_a^2/\vert M_3\vert^2$.
In this contribution the related CP phases are $\phi_{A_\alpha}$ and
$\phi_\mu$ in the squark mixing factors. This contribution can be small
enough if the off-diagonal element of the squark mass matrix 
$\vert A_\alpha\vert e^{i\phi_{A_\alpha}}
+\vert\mu\vert e^{-i\mu}R_f$ cancels by itself.

On the chargino and neutralino contributions to the quark EDM, we can
calculate it as in the same way as the electron case.
We find that it can be written as
\begin{eqnarray}
&&d_u^\chi/e={-\alpha\over 8\pi s_W^2}m_u\left[{1\over m_u}\sum_{j,a}
{2\over 3m_j}G(x_{aj}){\rm Im}({\cal A}^{u_a}_{\chi^0_j}) \right. \nonumber \\
&&\hspace*{3cm}\left.+{1\over m_W}\sum_{j,a}{1\over m_j}
\left\{F(x_{aj}- {1\over 3}G(x_{aj})\right\}{\rm Im}
({\cal A}_{\chi^\pm_j}^{u_a})\right], \nonumber \\
&&d_d^\chi/e={-\alpha\over 8\pi s_W^2}m_d\left[{1\over m_d}\sum_{j,a}
{-1\over 3m_j}G(x_{aj}){\rm Im}({\cal A}^{d_a}_{\chi^0_j}) \right. \nonumber \\
&&\hspace*{3cm}\left.+{1\over m_W}\sum_{j,a}{1\over m_j}
\left\{-F(x_{aj}) +{2\over 3}G(x_{aj})\right\}{\rm Im}
({\cal A}_{\chi^\pm_j}^{d_a})\right],
\end{eqnarray} 
where the mixing factors ${\cal A}^f_{\chi^\pm}$ and 
${\cal A}^f_{\chi^0}$ are defined as
\begin{eqnarray}
&&{\cal A}^{u_a}_{\chi^\pm_j}={1\over\sqrt{2}\sin\beta}W^{(+)}_{1j}W^{(-)}_{2j}
\vert V^d_{1a}\vert^2 +{1\over 2\sin\beta\cos\beta}{m_d\over m_W}
W^{(+)}_{1j}W^{(-)}_{1j}V^{d\ast}_{2a}V^d_{1a}, \nonumber \\
&&{\cal A}^{d_a}_{\chi^\pm_j}={1\over\sqrt{2}\cos\beta}W^{(-)}_{1j}W^{(+)}_{2j}
\vert V^u_{1a}\vert^2 +{1\over 2\sin\beta\cos\beta}{m_u\over m_W}
W^{(+)}_{1j}W^{(-)}_{1j}V^{u\ast}_{2a}V^u_{1a}, \nonumber \\
&&{\cal A}_{\chi^0_j}^{u_a}=-\left[\left({2\over 9}t_W^2U_{1j}^2
+{2\over 3}t_WU_{1j}U_{2j}\right)V_{1a}^{u\ast}V_{2a}^u\right. \nonumber \\
&&\hspace*{1cm}\left. -{m_u\over 2m_W\sin\beta}\left\{
\left({1\over 3}t_WU_{1j}U_{4j}+U_{2j}U_{4j}\right)\vert V_{1a}^u\vert^2
-{4\over 3}t_WU_{1j}U_{4j}\vert V_{2a}^u\vert^2\right\}\right]
\nonumber \\
&&{\cal A}_{\chi^0_j}^{d_a}=-\left[-\left({1\over 9}
t_W^2U_{1j}^2 -{1\over 3}t_W U_{1j}U_{2j}\right)V_{1a}^{d\ast}V_{2a}^d\right.
\nonumber \\
&&\hspace*{1cm}\left.-{m_d\over 2m_W\cos\beta}\left\{\left(
{1\over 3}t_WU_{1j}U_{3j}-U_{2j}U_{3j}\right)\vert V_{1a}^d\vert^2
+{2\over 3}t_WU_{1j}U_{3j}\vert V_{2a}^d\vert^2\right\}\right],
\end{eqnarray}
where we again neglect the higher order terms of the quark mass
in ${\cal A}^f_{\chi^0}$.
The situation for the cancellation among these contributions 
is just the same as the electron case. The statement in the footnote~1
should be also reminded for the cancellation in the neutron EDM.

\section{Numerical analysis}
In this section we present the results of the numerical calculation of 
the masses of the superpartners, the EDM of the electron and the neutron 
and the AMM of the muon in our model.
Before discussing the results of the caluculation 
we briefly explain our procedure for the numerical calculation.

We use the soft supersymmetry breaking parameters obtained in sec.~2 as the
initial values at a suitable supersymmetry breaking scale and make them evolve 
to the weak scale by using the one-loop renormalization group equations
(RGEs), except for the gauge and Yukawa couplings for which we use
two-loop RGEs.
Using the low energy parameters obtained in this way, we calculate the 
EDM of the electron and the neutron.

Free parameters related to the supersymmetry breaking
are $\Lambda_1$ and $\Lambda_2$, which 
determine all of the masses of the gauginos and the scalar superpartners.  
Only the difference of their phases $\theta_1$ and $\theta_2$ is
independent and it is related to the physical phases $\phi_3$ and 
$\phi_1$ defined in eq.~(\ref{eqii}) through eq.~(\ref{eqfff}).
Since $\mu$, $B$, and $A_\alpha$ are dependent on the model and cannot
be restricted in the present framework as stressed
in the previous part, we do not fix their origin and we treat them 
as free parameters.\footnote{If we
assume $A_\alpha(\Lambda)=B(\Lambda)=0$ in eq.~(\ref{eqg}), they can 
be definitely determined through the radiative effect by using $\Lambda_{1,2}$.
However, we do not adopt this possibility but treat them in 
more general way. }
If we assume the universality of $A_\alpha(\Lambda)$ such as 
$A_\alpha(\Lambda)=A$,
there are five independent real parameters 
$\phi_\mu$, $\phi_A$, $\vert\mu\vert$, $\vert B\vert$, and $\vert
A\vert$ where $\phi_B$ can be related to $\phi_\mu$ by imposing the
Higgs VEVs to be real under a phase convention such that $B\mu$ is real. 
Thus the soft supersymmetry breaking parameters are 
totally composed of 8 free real parameters in this study:
\begin{equation}
\vert\Lambda_1\vert, \quad \vert\Lambda_2\vert, \quad
\vert A\vert , \quad \vert B\vert, \quad \vert\mu\vert,\quad
\tilde\theta, \quad \phi_{A}, \quad \phi_\mu, 
\label{eqrr}
\end{equation}
where we define $\tilde\theta=\theta_1-\theta_2$.
There is an ambiguity on the scale where the soft supersymmetry breaking
parameters are introduced and start running. In the present analysis  
we take $\Lambda={\rm max}~(\vert\Lambda_1\vert, \vert\Lambda_2\vert)$
as such a scale, for simplicity.
This prescription is not expected to affect the results
largely.\footnote{Since we mainly study the region where
$\vert\Lambda_2\vert/\vert\Lambda_1\vert$ is not so large, this
treatment will be justified.}

In the region from the gauge coupling unification scale $M_{\rm U}$ 
to $\Lambda$ the RGEs for the gauge coupling constants and Yukawa coupling
constants are composed of the supersymmetric ones. The $\beta$-functions 
are calculated for the MSSM contents
and the messenger fields. We solve these for various initial values 
of the Yukawa couplings. At $\Lambda$ the messenger fields are supposed
to decouple and the soft supersymmetry breaking
parameters are introduced. Thus the RGEs become the same as the ones
of the MSSM. We can obtain their values at the electroweak scale 
by solving these RGEs numerically. 
In order to determine the phenomenologically interesting parameter region,
we impose several conditions on the weak scale parameters obtained by the RGEs.
As such conditions we adopt the following ones:\\
(i) The physical true vacuum should be 
radiatively realized as the minimum of the tree-level scalar potential 
in the consistent way with the masses of the top, bottom quarks and the tau
lepton. We check the consistency between the values of $\tan\beta$
predicted from these two different physical requirement.\\
(ii) Various experimental mass bounds for the superpartners, such as 
gluinos, charginos, stop, stau, and the charged Higgs scalar 
should be satisfied.\footnote{In this study the large CP-phases are
assumed to exist. Since the CP even neutral Higgs scalars mix with the
CP odd neutral Higgs scalar, the Higgs mass formulas are changed
\cite{cphiggs}.  Thus we do not impose the neutral Higgs mass bound.}
 The color and electromagnetic charge should not be broken.\\ 
After restricting the parameter space at the high energy scale by imposing 
these conditions on the weak scale values, we finally calculate 
the EDM of the electron and the neutron and impose their 
experimental bounds  \cite{eedm,nedm}
$$
\vert d_{e}/e \vert=1.6\times 10^{-27}~{\rm cm},
\qquad  
\vert d_{n}/e \vert =1.2\times 10^{-25} {\rm cm},
$$
to restrict the selected parameter region further.

In the following part we focus our stusy into the cases with the large
physical CP-phases such as $\phi_A=\pi/2$ and $ \tilde\theta= 3\pi/2$.
Although we search the possible region of $\Lambda_{1,2}$ and $\phi_\mu$,  
we restrict our study for the parameters $\vert A\vert,~\vert B\vert$ and 
$\vert \mu\vert$ into
$100~{\rm GeV}\le \vert A\vert,~\vert B\vert,~\vert \mu\vert 
\le 500~{\rm GeV}$.\footnote{We also study the case of $A=0$, in which 
the free parameter can be reduced into six. The result is not much
different from this case and it can be included in the region given here.}

\subsection{Spectrum of superpartners}
In the present model the mass parameters of the superpartners are represented 
by the restricted number of input parameters as shown in the previous 
part. Their masses are all determined only by $\Lambda_{1,2}$. 
The mass parameters of the superpartners are also related to 
the realization of the radiative symmetry breaking at the weak scale.
This fact makes intimately relate $\mu$ and $B$ to $\Lambda_{1,2}$
in the present model.  
The minimum conditions for the tree-level scalar potential 
can be written as
\begin{equation} 
\sin 2\beta={2B\mu \over m_1^2 +m_2^2 +2\vert\mu\vert^2}, \qquad
m_Z^2=-2\vert\mu\vert^2+{2m_1^2-2m_2^2\tan^2\beta\over \tan^2\beta -1}.
\end{equation}
These conditions tell us what kind of tuning for the $\mu$ and $B$ parameters 
are required to realize the correct vacuum in the present supersymmetry 
breaking scenario \cite{mu}. Since we have no knowledge for $\mu$ and
$B$ in the
present scenario, the study of these conditions seems to give us a useful
information for them.
In Fig.~1 we show what kind of values of $\mu$ and $B$
are required as a function of $\vert\Lambda_2\vert/\vert\Lambda_1\vert$. 
This figure shows that we can have the most solutions in the region of $\vert
B\vert>\vert\mu\vert$ for $\vert\Lambda_2\vert/\vert\Lambda_1
\vert~{^>_\sim}~1.7$,
although there are solutions only in the region of 
$\vert B\vert<\vert\mu\vert$ for
$\vert\Lambda_2\vert/\vert\Lambda_1\vert\simeq 1$
which corresponds to the ordinary minimal GMSB \cite{sue1}.
This shows that the present model can relax the condition 
for the relative magnitude of $\vert\mu\vert$ and $\vert B\vert$ 
which is required by the radiative symmetry breaking. 
Since the value of $\tan\beta$ tends to be small in the case of 
$\vert\Lambda_2\vert>\vert\Lambda_1\vert$ as discussed in \cite{sue1},
we restrict the $\tan\beta$ value into the favored region such as 
$2.7~{^<_\sim}~\tan\beta~{^<_\sim}~3.2$ and present the results.

\input epsf 
\begin{figure}[tb]
\begin{center}
\epsfxsize=7.4cm
\leavevmode
\epsfbox{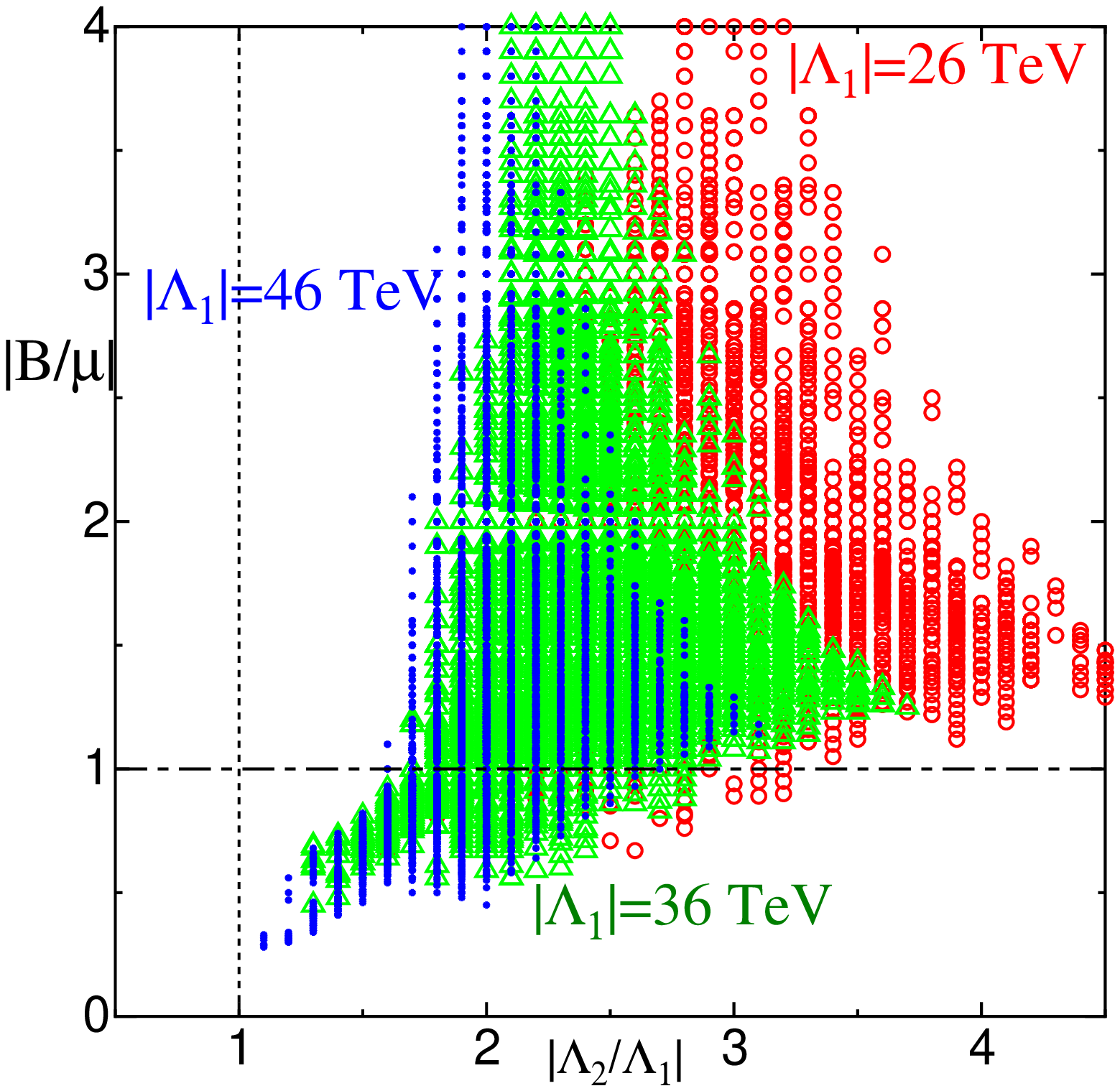}\\
\end{center}
\vspace*{-0.5cm}
{\footnotesize Fig. 1~~
The ratio of $\vert B\vert$ to $\vert\mu\vert$
 required by the radiative symmetry breaking conditions.}
\end{figure}

As stressed before, the mass formulas of the superpartners can be 
written only by two parameters $\Lambda_{1,2}$ and then the model is 
very predictive at least for the masses of superpartners.
It is interesting that the spectrum can be largely different from the
ones of the ordinary GMSB scenario. 
In order to display the feature of the mass spectrum of superpartners, 
it is convenient to plot them as the functions of
$\vert\Lambda_2\vert/\vert\Lambda_1\vert$. We give them in Fig.~2.
As is easily seen from these figures, this scenario predicts that 
the next lightest superparticle (NLSP) can be a neutralino.
This is rather different feature 
from the ordinary GMSB where the NLSP tends to be the right-handed stau. 
The right-handed stop is rather light in the 
$\vert\Lambda_2\vert/\vert\Lambda_1\vert>1$ region. Since the 
right-handed stop has the contribution only through the U(1)$_Y$
coupling at $\vert\Lambda_2\vert$ and also the gluino mass which 
is determined by $\vert\Lambda_1\vert$ is small, 
the RGE evolution can reduce the 
right-handed stop mass in the case of the large 
$\vert\Lambda_2\vert/\vert\Lambda_1\vert$. 
These are related to the general feature of this model such that 
the SU(2) nonsinglet fields tend to be heavier than the 
SU(2) singlet fields. This type of spectrum of superpartners can 
make the unification scale of the gauge coupling constants higher than
the ordinary one of the MSSM.\footnote{This possibility has been 
discussed in the different context in \cite{coup}. Our present model can
realize such a spectrum in a natural way.}

\begin{figure}[tb]
\begin{center}
\epsfxsize=7.4cm
\leavevmode
\epsfbox{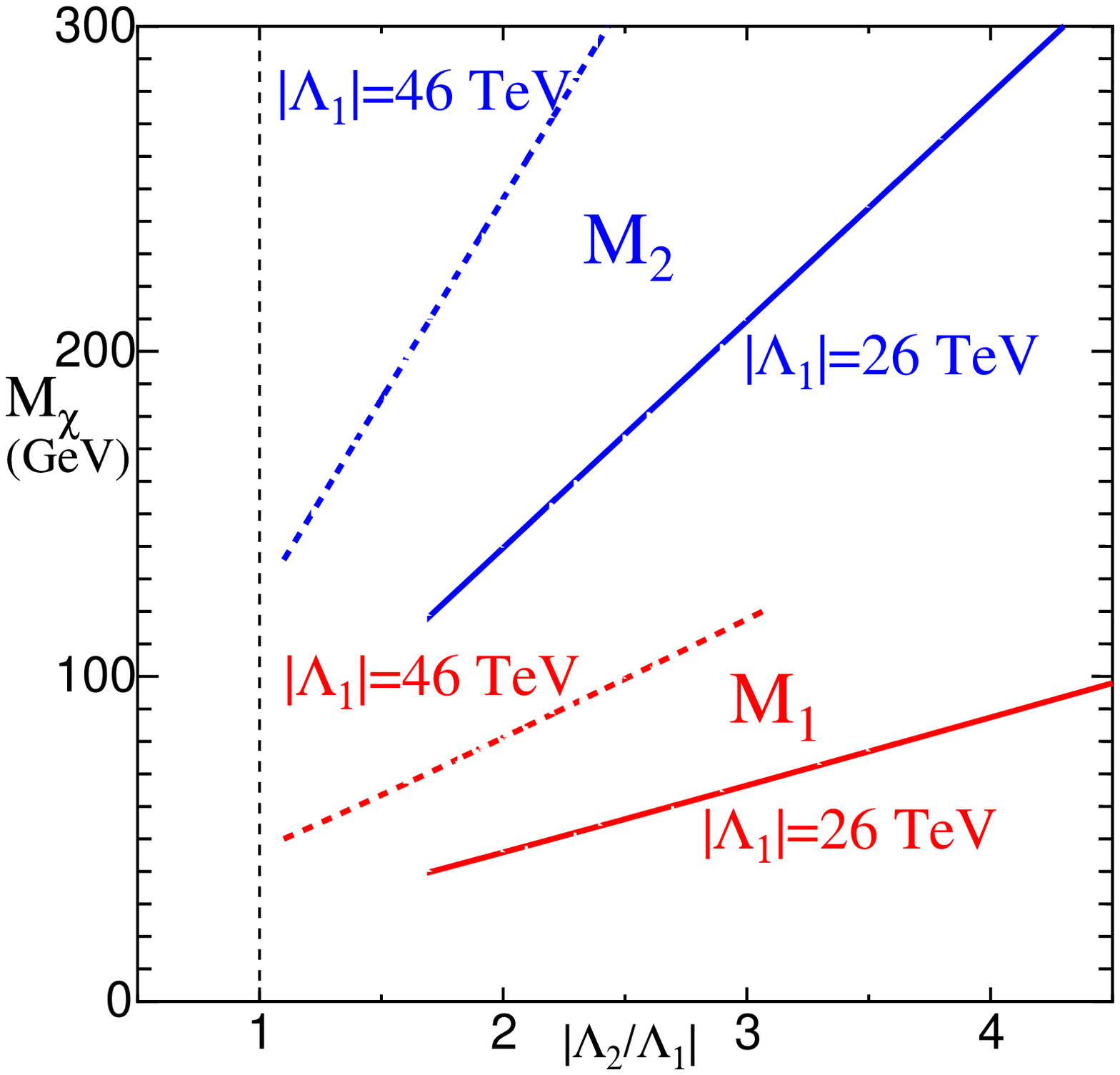}
\hspace*{0.3cm}
\epsfxsize=7.4cm
\leavevmode
\epsfbox{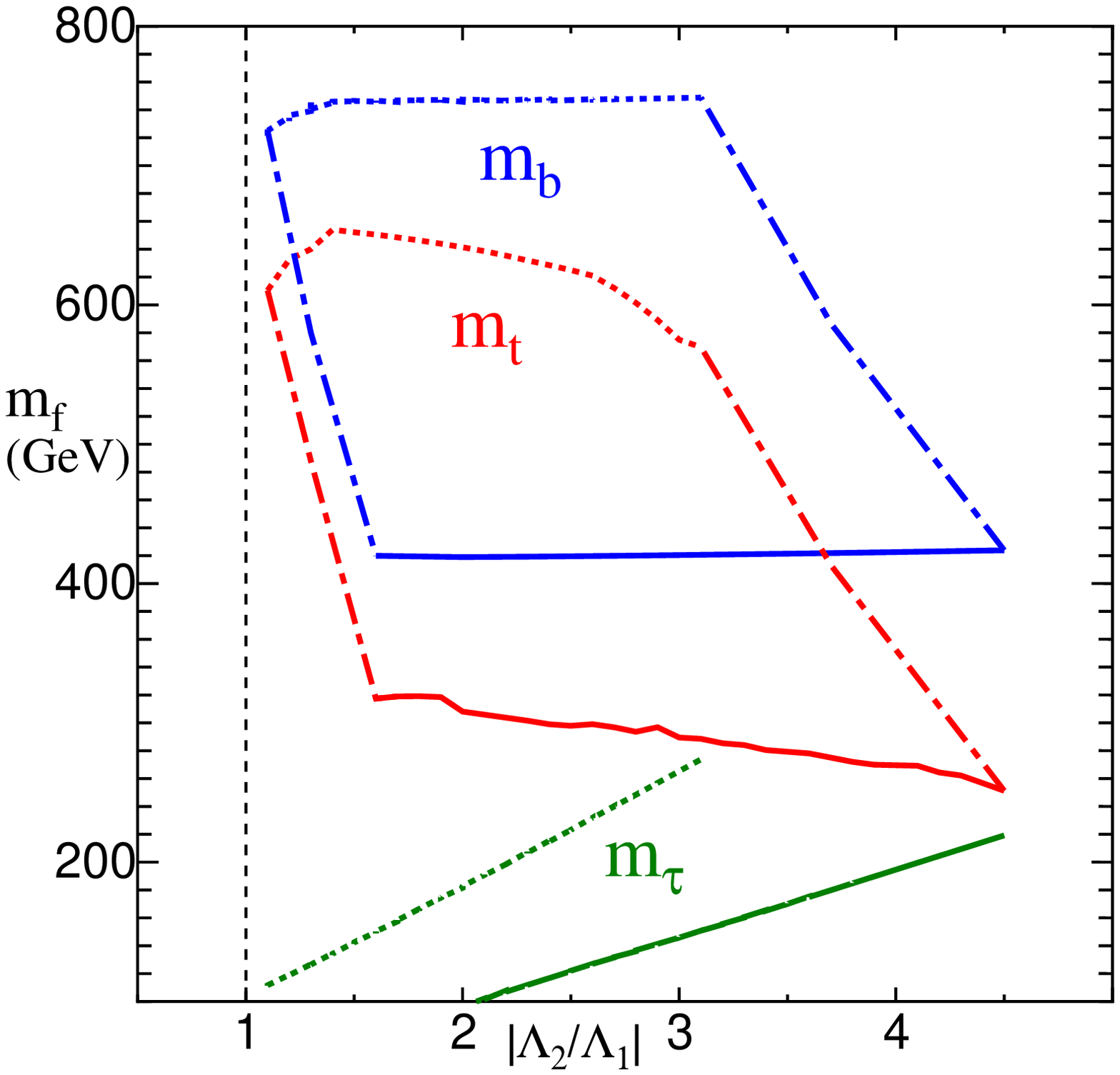}\\
\end{center}
\vspace*{-4mm}
{\footnotesize Fig. 2~~\  Mass spectrum of the low lying superpartners
for various values of $\vert\Lambda_{1,2}\vert$.
The left panel shows the gaugino masses related to charginos and 
neutralinos. The right panel shows the third generation sfermion masses.
In each panel the solid lines correpsond to
 $\vert\Lambda_1\vert=26~{\rm TeV}$ and the dotted lines correspond to 
$\vert\Lambda_1\vert=46~{\rm TeV}$.}
\end{figure}

\subsection{EDMs of the electron and the neutron}
In the previous section we have discussed what kind of contributions 
for the EDM can exist as the effects of the superpartners.
Here we show two important ingredients for 
the cancellations between various contributions to the EDM by using the
numerical analysis.
In the left panel of Fig.~3 we plot the allowed upper and lower bound values 
of $\vert\Lambda_2\vert$ as a function of $\vert\Lambda_1\vert$ 
by imposing the phenomenological constraints.  
This shows that we can obtain the solutions only in the 
$\vert\Lambda_1\vert<\vert\Lambda_2\vert$ region. 
If we do not impose the EDM constraints, we can find the
solutions also in the region $\vert\Lambda_1\vert>\vert\Lambda_2\vert$ 
\cite{sue1}.
Thus this result is considered to be caused by the EDM constraints. 
In this region the neutralino can be much lighter than the chargino,
which is expected from the superpartners mass formulas. (See Fig.~2
also.) This aspect of the mass spectrum seems to make the neutralino
contribution larger and then the cancellations
between the chargino contribution and the neutralino contribution to the 
EDM is considered to be effective. 
\begin{figure}[tb]
\begin{center}
\epsfxsize=7.4cm
\leavevmode
\epsfbox{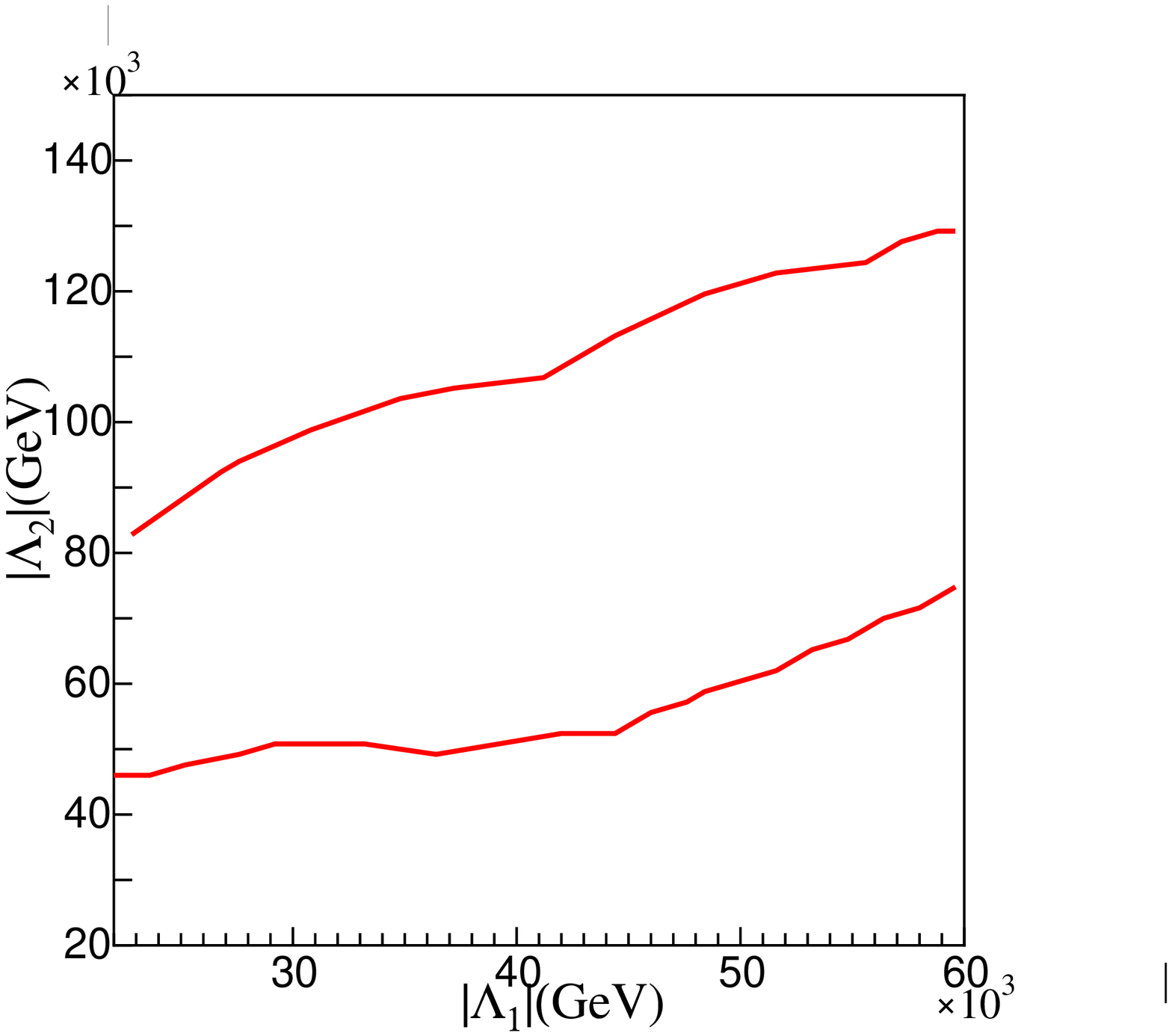}
\hspace*{0.3cm}
\epsfxsize=7.4cm
\leavevmode
\epsfbox{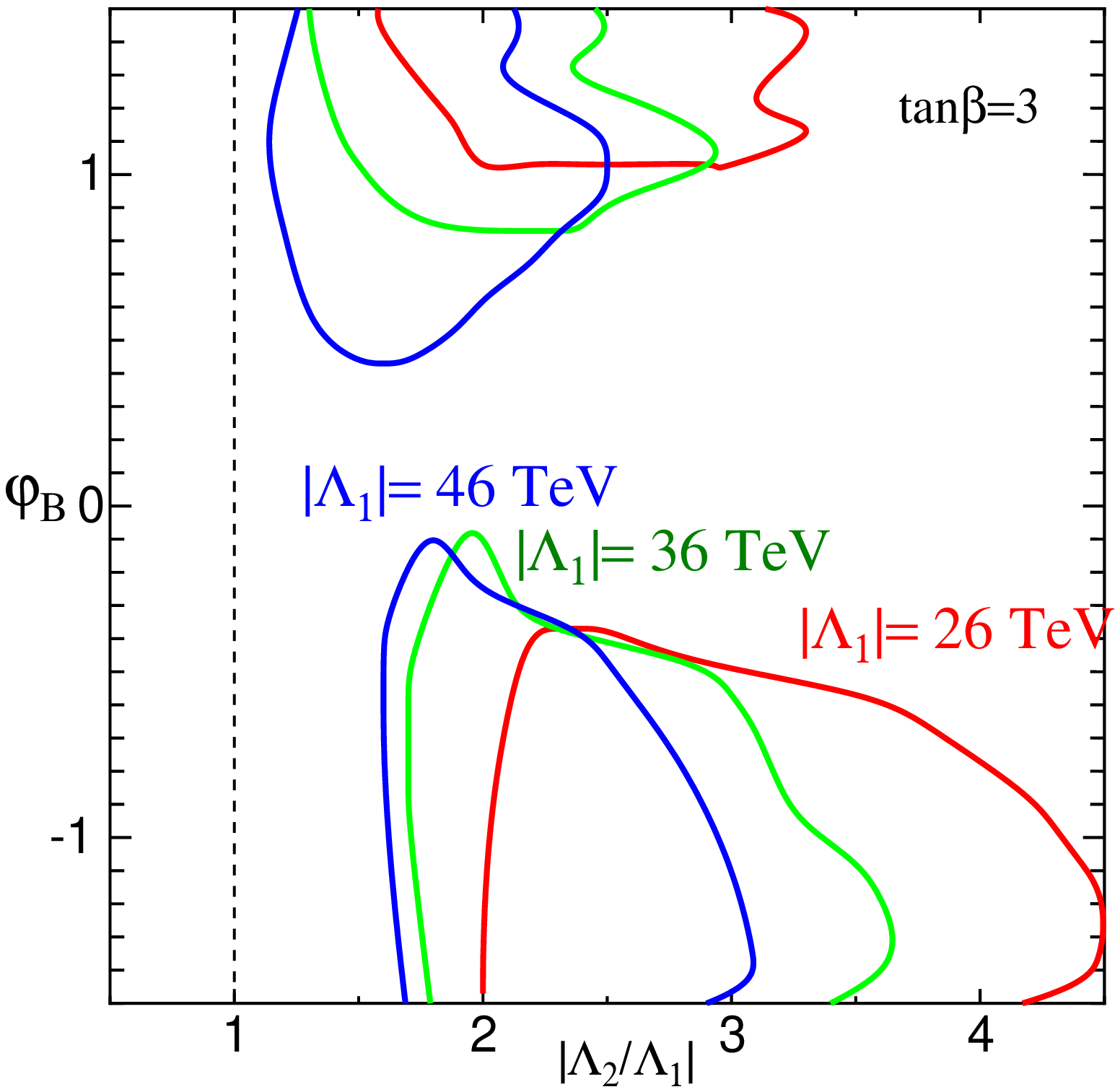}\\
\vspace*{-0.3cm}
\end{center}
{\footnotesize Fig. 3~~\  
The left panel shows the allowed region in the 
$(\vert\Lambda_1\vert,~\vert\Lambda_2\vert)$ plane
which satisfies both the EDM constraints and the phenomenological 
conditions given in the text.
The right panel shows the allowed region of $\phi_B(= -\phi_\mu)$ 
for the same constraints. In both panel the region surrounded by the
 curves are allowed.}
\end{figure}

On the CP-phases we have discussed how each phase contributes to the EDM
in the previous section.
In this numerical study we assume the maximum values for $\phi_A$ and
$\tilde\theta$. The large $\tilde\theta$ results in the large CP-phases in the
gaugino sector. In the right panel of Fig.~3 we show the required value
for $\phi_\mu$ to satisfy the EDM constraints. We can
find that the rather large value of $\phi_\mu$ is required for the
cancellations of the various contributions to the EDM of the electron
and the neutron, as is expected
from the previous discussion.  
This result shows that the large CP-phases in the soft supersymmetry
breaking parameters can be consistent with the EDM constraints of the
electron and the neutron as far as the large CP-phases exist in the
gaugino sector. 

It is interesting that our model can realize the 
convenient situation for the cancellation for the EDM of the electron
and the neutron, that is,
the desirable mass spectrum for the case of
$\vert\Lambda_2\vert>\vert\Lambda_1\vert$ and also the presence of physical 
CP-phases in the gaugino sector. 
The predicted lower bounds for the electron EDM and the neutron EDM are
$$\vert d_e/e\vert~{^>_\sim}~10^{-31\sim -30}~{\rm cm}, \qquad
\vert d_n/e\vert~{^>_\sim}~10^{-29\sim -28}~{\rm cm}.$$
We also show the predicted value for the AMM of the muon in Fig.~4.
These values are less than the half of the value of 
the difference between the experimental value and the SM prediction,
$a_\mu^{\rm exp}-a_\mu^{\rm SM}=(345\pm 114)\times 10^{-11}$,
which is presented in \cite{amm}. 

\begin{figure}[tb]
\begin{center}
\epsfxsize=7.4cm
\leavevmode
\epsfbox{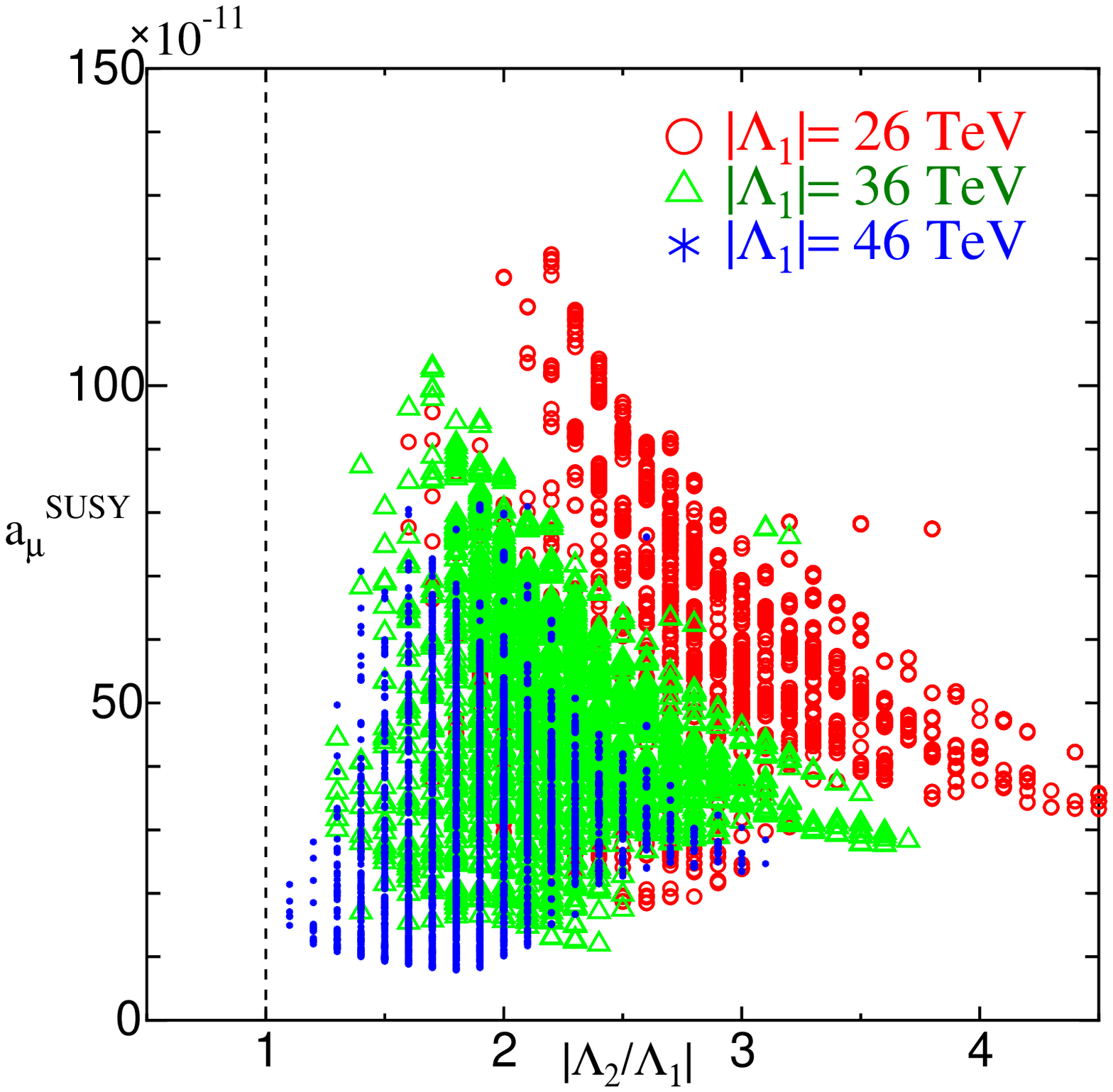}\\
\vspace*{-0.3cm}
\end{center}
{\footnotesize Fig. 4~~\  The predicted value of the 
AMM of a muon in the parameter space where the EDM bounds and the
 phenomenological constraints are satisfied.}
\end{figure}

\section{Summary}
We have investigated the non-universal CP-phases in the gaugino masses 
and their effects on the EDM constraints in the extended GMSB. 
We have shown that the non-universal gaugino masses can be generally 
realized if we assume that the SU(3) triplet and the SU(2) 
doublet messenger fields couple to the different singlet chiral 
superfields which are assumed to break the supersymmetry through the 
hidden sector dynamics.
As such an example we have presented a model with the direct product 
gauge structure SU(5)$^\prime\times$SU(5)$^{\prime\prime}$ in the appendix. 
In this model the discrete symmetry introduced to realize the 
doublet-triplet splitting simultaneously forces the SU(3) triplet 
and the SU(2) doublet messenger fields to couple to the different 
singlet chiral superfields.

In this type of model the characteristic spectrum of superpartners 
is induced and also the non-universal CP-phases in the gaugino masses 
can be introduced. 
The SU(2) nonsinglet superpartners tend to be heavier than the
SU(2) singlet ones. The CP-phases can remain in the gaugino sector 
as the physical ones after the $R$-transformation. 
These may result in the various interesting phenomenology different from 
ordinary GMSB scenario.
As the most interesting one, we have calculated the effect of the 
CP-phases on the EDM of the electron and the neutron by solving the 
RGEs for the soft supersymmetry breaking parameters obtained in our
model. We have found that the experimental bounds can be satisfied since 
the effective cancellation occurs between the neutralino 
and chargino contributions even for the order one CP-phases without 
assuming the heavy superpartners of the $O(1)$~TeV masses.
This cancellation is considered to be caused mainly by the existence 
of the CP-phases in the
gaugino sector and the feature of the mass spectrum such that the
neutralino can be much lighter than the chargino.
The same origin for them also makes the right-handed stop rather light 
and also the neutralino lighter than the right-handed stau. 
Since the SU(2) nonsinglet superpartners may decouple earlier than 
others, the gauge coupling unification can be realized at the higher 
scale than the MSSM. 

Further phenomenological study of the model seems to be necessary since
the essential feature of the model may be related to the reasonable
motivation to solve the doublet-triplet splitting problem \cite{dt} in the 
grand unified model.
In particular, the Higgs sector can be affected by the the existence of
the large CP-phases in the soft supersymmetry breaking parameters
\cite{higgs}.
Since the CP even Higgs scalar field can be mixed with the CP odd Higgs 
scalar, the lightest neutral Higgs mass can be largely modified from the 
one of the ordinary case. This aspect of the present model will be 
studied elsewhere.

\vspace{.5cm}
\noindent
This work is supported in part by a Grant-in-Aid for Scientific 
Research (C) from Japan Society for Promotion of Science
(No.~14540251) and also by a Grant-in-Aid for Scientific 
Research on Priority Areas (A) from The Ministry of Education, Science,
Sports and Culture (No.~14039205).

\newpage
\noindent
{\Large\bf Appendix}

In this appendix we give an example which can realize the extended 
GMSB \cite{sue}.
The model is defined by the direct product gauge structure such as 
${\cal G}=$SU(5)$^\prime\times$SU(5)$^{\prime\prime}$ and a
global discrete symmetry $F$ which commutes with this gauge 
symmetry ${\cal G}$.
We introduce the chiral superfields summarized in Table 1.
In order to cause the symmetry breaking at the high energy scale
the ${\cal G}\times F$ invariant renormalizable superpotential 
for $\Sigma$, $\Phi_1$ and $\Phi_2$ is assumed as
\begin{equation}
W_1= M_\phi~{\rm Tr}\left(\Phi_1\Phi_2\right) 
+{1\over 2}M_\sigma~{\rm Tr}\left(\Sigma^2\right)
+\lambda~{\rm Tr}\left(\Phi_1\Sigma\Phi_2 +{1\over 3}\Sigma^3\right), 
\end{equation} 
The scalar potential induced from this $W_1$ can be obtained as
\begin{eqnarray}
&&V={\rm Tr}~\vert M\phi_1+\lambda\phi_1\sigma + y\vert^2
+ {\rm Tr}~\vert M\phi_2+\lambda\sigma\phi_2 + x \vert^2 \nonumber \\
&&\hspace*{0.5cm} +{\rm Tr}~\vert M\sigma 
+\lambda\phi_1\phi_2+\sigma^2 +z \vert^2, 
\end{eqnarray}
where $\phi_{1,2}$ and $\sigma$ are the scalar components of
$\Phi_{1,2}$ and $\Sigma$, respectively. 
They are traceless and $x$, $y$ and  $z$ are the Lagrange multipliers 
for these traceless conditions.
We can easily find a non-trivial solution of the $V$ minimum such as
\begin{eqnarray}
&&\phi_2={x\over y}\phi_1, \label{eqa1}\\
&&M_\phi\phi_1 +\lambda\phi_1\sigma+y=0, \label{eqa2} \\
&&M_\sigma\sigma +\lambda\left(\sigma^2+{x\over y}\phi_1^2\right)+z=0, 
\label{eqa3}
\end{eqnarray}
where the Lagrange multipliers $y$ and $z$ are determined as
\begin{equation}
y=-{\lambda\over 5}{\rm Tr}\left(\phi_1\sigma\right), \qquad
z=-{\lambda\over 5}{\rm Tr}\left(\sigma^2
-{5x\over \lambda{\rm Tr}(\phi_1\sigma)}\right),
\end{equation}
and $x$ remains as a free parameter.
If we restrict ourselves to a special direction in the field space
such as $\phi_1=\kappa\sigma$ and also assume 
$M_\sigma=M_\phi(1+x\kappa^2/y)$,
eqs.~(\ref{eqa2}) and (\ref{eqa3}) are reduced into the same 
equation for the adjoint Higgs scalar in the ordinary supersymmetric SU(5)
model as
\begin{equation}
M_\phi\sigma +\lambda\sigma^2 -{\lambda\over 5}{\rm Tr}
\left(\sigma^2\right)=0.
\end{equation} 
We adopt the most interesting one among three degenerate independent 
solutions, which can be written as 
\begin{equation}
\sigma=\tilde M~{\rm diag}~(2,~2,~2,~-3,~-3),
\label{eqa}
\end{equation}
where $\tilde M$ is defined as $\tilde M=M_\phi/\lambda$.
Using this $\sigma$, other fields are determined as
\begin{equation}
\phi_1=\kappa\sigma, \qquad \phi_2={1\over\kappa}
\left({M_\sigma\over M_\phi}-1\right)\sigma,
\label{eqa0}
\end{equation}
where $\kappa$ is an undetermined parameter.
\begin{figure}[tb]
\begin{center}
\begin{tabular}{|l|c|c|c|c|}\hline
         & ${\cal F}({\cal G}~{\rm rep.})$  &  $F$    
& \multicolumn{2}{|c|}{$F^\prime$} \\\cline{4-5}
   &&& ${\bf 3}\in {\bf 5}$ or $\bar{\bf 3}\in\bar{\bf 5}$ & 
${\bf 2}\in{\bf 5}$ or $\bar{\bf 2}\in\bar{\bf 5}$ \\\hline
{\rm Quarks/Leptons}& $\Psi^j_{10}({\bf 10}, {\bf 1})$ &  $\alpha$ & 
$\alpha$ & $\alpha$ \\
 $(j=1\sim 3)$ & $\Psi^j_{\bar 5}(\bar{\bf 5}, {\bf 1})$ &$\beta$ & $\beta$ 
& $\beta$ \\\hline 
{\rm Higgs fields}& $H({\bf 5}, {\bf 1})$ & $\gamma$ & $\gamma$ & $\gamma$\\
           & $\tilde H({\bf 1}, \bar{\bf 5})$ & $\xi$ 
& $\xi+2a$ & $\xi-3a$ \\ \hline
{\rm Messenger fields}& $\bar\chi(\bar{\bf 5}, {\bf 1})$ & $\delta$ & 
$\delta$ & $\delta$\\ 
              & $\chi({\bf 1}, {\bf 5})$ & $\zeta$ & 
$\zeta-2a$ & $\zeta+3a$
 \\ \hline
{\rm Bifundamental field}  & $\Phi_1(\bar{\bf 5},{\bf 5})$ & $\eta$ 
& $\eta+2b$ &  
$\eta-3b$ \\
            & $\Phi_2({\bf 5},\bar{\bf 5})$ & $\sigma$ & $\sigma-2b$ & 
$\sigma+3b$ \\\hline
{\rm Adjoint Higgs field}  & $\Sigma({\bf 1},{\bf 24})$ & 0 & 
\multicolumn{2}{|c|}{0\hspace{2mm}( {\rm for}~$\Sigma^{\bar i}_i$ )}     
\\\hline
{\rm Singlets}   & $S_1({\bf 1}, {\bf 1})$& $\theta$ &  
\multicolumn{2}{|c|}{$\theta$}\\
                 &$S_2({\bf 1}, {\bf 1})$ & $\tau$ & 
\multicolumn{2}{|c|}{$\tau$}\\
\hline
\end{tabular}
\vspace*{3mm}\\
\end{center}
{\footnotesize Table 1~ Discrete charge assignment for the chiral 
superfields. For the adjoint Higgs field $\Sigma$ we only give the charge for
 the diagonal components.} 
\end{figure} 

The vacuum defined by eqs.~(\ref{eqa}) and (\ref{eqa0})
is found to be invariant under the gauge transformation of 
${\cal H}$=SU(3)$\times$S(2)$\times$U(1) which is the subgroup of
the diagonal sum SU(5) of ${\cal G}$.
If we assume that the model is based on the deconstruction method \cite{de}, 
the remaining discrete symmetry is $F^\prime= F\times
G_{U(1)^{\prime\prime}}$ where $G_{U(1)^{\prime\prime}}$ is the discrete
subgroup of the hypercharge U(1)$^{\prime\prime}$ of 
SU(5)$^{\prime\prime}$ \cite{w}.
The charge assignment of $F^\prime$ is shown in Table~1.\footnote{We
assume that SU(5)$^{\prime\prime}$ is induced as the diagonal sum of two 
SU(5) effectively and also $\tilde H, \chi$ and $\Phi_1,~\Phi_2$ belong to the
different SU(5), respectively. This assumption makes it possible to
introduce the independent charge normalization $a$ and $b$ for
$G_{U(1)^{\prime\prime}}$. 
Since $G_{U(1)^{\prime\prime}}$ is not just U(1)$^{\prime\prime}$ but
its discrete subgroup $Z_n$, its charges of $\tilde H,~\chi$ and
$\Phi_{1,2}$ can be taken independently
and the assumption for SU(5)$^{\prime\prime}$ may not be ncessary.}
After the symmetry breaking from ${\cal G}\times F$ into ${\cal H}\times 
F^\prime$, superpotential structure is expected to be changed into 
the ${\cal H}\times F^\prime$ invariant one as in the case of the Wilson 
line breaking in the heterotic string.

We require various conditions on $F$ and $F^\prime$ to satisfy 
phenomenological constraints to realize our purpose. As such conditions
we take the following ones.\\
(i)~Each term in the $F$ invariant superpotential $W_1$ should exist 
before the symmetry breaking and this
requirement imposes the condition
\begin{equation}
\eta+\sigma =0.
\end{equation}
(ii)~The gauge invariant bare mass terms of the fields such as 
$\Psi_{\bar 5}H$, $H\bar\chi$, $\tilde H\chi$ and $S_\alpha S_\beta$
should be forbidden by both $F$ and $F^\prime$. 
These conditions are summarized as
\begin{eqnarray}
&&\beta+\gamma\not=0, \qquad \gamma+\delta\not=0,\qquad  
\xi+\zeta \not=0, \nonumber \\
&& 2\theta\not=0,\qquad 2\tau\not=0, \qquad \theta+\tau\not=0.
\end{eqnarray}
(iii)~To realize the doublet-triplet splitting \cite{dt}
Yukawa coupling $\Phi_1 H\tilde H$ should be forbidden by $F$.
Moreover, $\Phi_{1,2} H_2\tilde H_{2}$ and $\Sigma H_2\tilde H_{2}$ 
should also be forbidden by $F^\prime$ after the symmetry breaking, 
although $\Phi_1H_3\tilde H_{3}$ is allowed at least. 
This gives the conditions such as
\begin{eqnarray}
&&\gamma+\xi+\eta \not=0, \quad  \gamma+\xi+\eta+2(a+b)=0,\nonumber \\
&&\gamma+\xi+\eta-3(a+b)\not=0,\quad 
 \gamma+\xi+\sigma-3(a-b)\not=0, \quad \gamma+\xi-3a\not=0. 
\end{eqnarray}
(iv)~Yukawa couplings of quarks and leptons, that is, 
$\Psi_{10}\Psi_{10}H_2$ and $\Psi_{10}\Psi_{\bar 5}\tilde H_{\bar
2}\Phi_1$ should exist at least under $F^\prime$. This requires
\begin{equation}
 2\alpha + \gamma=0, \qquad 
\alpha+\beta+\xi+\eta-3(a+b)=0.
\end{equation} 
(v)~The chiral superfields $\chi$ and $\bar\chi$ should be massless
at the ${\cal G}$ breaking scale due to $F$ and they play the role 
of the messenger
fields of the supersymmetry breaking which is assumed to occur in the
$S_\alpha$ sector.
These require the absence of $\Phi_2\chi\bar\chi$ under $F$ 
and also the absence of $\Phi_{1,2}\chi\bar\chi$ and 
$\Sigma\chi\bar\chi$ under $F^\prime$, although 
the existence of $\Phi_2S_\alpha\chi\bar\chi$ under $F^\prime$ is needed. 
These conditions can be written as
\begin{eqnarray}
&&\delta+\zeta+\sigma\not=0, \quad 
\delta+\zeta+\sigma-2(a+b)+\theta=0,\quad \delta+\zeta+\sigma+3(a+b)+\tau=0,
\nonumber \\
&&\delta+\zeta-2a\not=0,\quad \delta+\zeta+\sigma-2(a+b)\not=0,\quad
\delta+\zeta+\eta-2(a-b)\not=0,\nonumber \\
&&\delta+\zeta+3a\not=0,\quad \delta+\zeta+\sigma+3(a+b)\not=0,\quad
\delta+\zeta+\eta+3(a-b)\not=0.
\label{mess}
\end{eqnarray}
(vi)~The neutrino should be massive and the proton should be
stable.\footnote{The magnitude of the neutrino masses realized in this
way depends on the details of the model and we do not
discuss this point further here.}
This means that $\Phi_{\bar 5}^2H_2^2$ should exist and
$\Psi_{10}\Psi_{\bar 5}^2$ and $\Psi_{10}^3\Psi_{\bar 5}$ should be
forbidden. These require 
\begin{equation}
2(\beta+\gamma)=0, \qquad \alpha+ 2\beta\not=0, \qquad
3\alpha+ \beta\not= 0.
\end{equation}
All of these conditions should be understood up to the modulus $n$ when we take
$F^\prime=Z_n$.

We can easily find an example of the consistent solution 
for these constraints. 
For example, if we take $F^\prime=Z_{20}$, such an example can be
given as
\begin{eqnarray}
&&\alpha=\eta=\zeta=-\sigma=a=1,\qquad \gamma=-b=-2, \nonumber \\ 
&&\delta=\theta=3, \qquad \xi=-5, \qquad \beta=-\tau=-8.
\end{eqnarray}
It should be noted that the existence of the different singlet 
fields $S_{1,2}$ is
generally required in order to make $\chi$ and $\bar\chi$ 
play a role of messengers of the supersymmetry breaking. 
In fact, the $F^\prime$ charges of $\chi$ and $\bar\chi$ satisfy
\begin{equation}
\theta-\tau=5(a+b)\not=0,    \qquad ({\rm mod}~n)
\end{equation}
which is derived from eq.~(\ref{mess}). This feature is caused by the
discrete symmetry which is related to the doublet-triplet splitting.


\end{document}